# Prioritizing High-Consequence Biological Capabilities in Evaluations of Artificial Intelligence Models

**Jaspreet Pannu[1,2], Doni Bloomfield[1], Alex Zhu[1], Robert MacKnight[3], Gabe Gomes[3,4,5], Anita Cicero[1], Thomas V. Inglesby[1]**

[1]Center for Health Security, Bloomberg School of Public Health, Johns Hopkins University [2]Department of Health Policy, Stanford School of Medicine, Stanford University [3]Department of Chemical Engineering, Carnegie Mellon University [4]Department of Chemistry, Carnegie Mellon University [5]Wilton E. Scott Institute for Energy Innovation, Carnegie Mellon University

## Abstract

As a result of rapidly accelerating artificial intelligence (AI) capabilities, over the past year, multiple national governments and multinational bodies have announced efforts to address safety, security and ethics issues related to AI models. One high priority among these efforts is the mitigation of misuse of AI models, such as for the development of chemical, biological, nuclear or radiological (CBRN) threats. Many biologists have for decades sought to reduce the risks of scientific research that could lead, through accident or misuse, to high-consequence disease outbreaks. Scientists have carefully considered what types of life sciences research have the potential for both benefit and risk (dual-use), especially as scientific advances have accelerated our ability to engineer organisms and create novel variants of pathogens. Here we describe how previous experience and study by scientists and policy professionals of dual-use capabilities in the life sciences can inform risk evaluations of AI models with biological capabilities. We argue that AI model evaluations should prioritize addressing high-consequence risks (those that could cause large-scale harm to the public, such as pandemics), and that these risks should be evaluated prior to model deployment so as to allow potential biosafety and/or biosecurity measures. While biological research is on balance immensely beneficial, it is well recognized that some biological research information and technologies could be intentionally or inadvertently misused to cause large-scale harm to the public. AI-enabled life sciences research is no different. Scientists' historical experience with identifying and mitigating dual-use biological risks can thus help inform new approaches to evaluating biological AI models. Identifying which AI capabilities pose the greatest biosecurity and biosafety concerns is necessary in order to establish targeted AI safety evaluation methods, secure these tools against accident and misuse, and avoid impeding immense potential benefits.

Correspondence to: tinglesby@jhu.edu

# Introduction

Scientists can perform almost all biological research in ways that pose minimal risk to society. However, some limited areas of life sciences research can threaten high-consequence harms to the public, through laboratory accidents or misuse. These risks are exacerbated by our growing capability to engineer existing pathogens and potentially create new ones—such as novel pandemic-capable variants or de novo synthesis of extinct pandemic pathogens that are no longer found outside of labs or repositories. Researchers can now also combine rapidly improving artificial intelligence (AI) models with wet-lab advances to facilitate, accelerate and augment this work.

Within the biological sciences, AI models will likely provide immense benefit. They are likely to be employed to improve the diagnosis and treatment of diseases, boost agricultural yields, and optimize the biosynthesis of useful products, among many other uses now being explored. Such models can make use of and generate biological information in the form of natural language, genetic sequences, protein sequences, protein structure, or even more sophisticated biological complexes and genetic regulatory architecture (1). AI models thus increase access to both basic biological information as well as more complex biomolecular designs and system behaviors. Indeed, biological AI models have already surpassed human performance on multiple tasks, and their development is advancing rapidly (2). However, high quality biological data remains a barrier to capability advancement.

As with all tools that allow us to better manipulate biology, future advanced AI models have the potential to be misused or misapplied, and these biosecurity risks have been publicly noted by scientists and model developers. Baker and Church note that AI protein design models are "vulnerable to misuse and the production of dangerous biological agents" while Boiko et al. include a security supplement in which they state that they "strongly believe that guardrails must be put in place to prevent … dual-use of large language models" for autonomous completion of chemical and biological synthesis protocols (3,4). Two authors of this paper were co-authors on Boiko et al.

> **Box 1. Recent biosecurity relevant trends in AI development**
>
> Concerns regarding potential biosecurity risks have been bolstered by recent trends in AI development.
>
> 1. *Large-language models* (LLMs) such as GPT-4 have shown rapid progress in dual-use capabilities, including assisting with completing biological and chemical research design and testing.
>
> 2. AI systems which include biological information, data and outputs, which we term *biological AI models*, have seen a similar rate of progress and model size expansion. This progress suggests the possibility that *biological foundation models* could be built: large, multi-purpose models that capture the complexity of biology.
>
> 3. Advances in LLMs and biological AI models are complementary. There are initial indications that LLMs can now assist users in accessing biological AI models to perform complex scientific tasks. These advances are in the future likely to lower the cost of achieving biological breakthroughs and allow less experienced researchers to use increasingly complex and powerful biological tools.
>
> 4. LLMs, biological AI models, and integrations of the two, can also interface with *AI-enabled autonomous laboratory environments*. These capabilities further reduce the time, expertise, and equipment required to synthesize pathogens, and suggest the possibility of end-to-end or "full stack" AI tool development in this domain.

In considering the biosecurity and biosafety implications of biological AI models, it is important to consider the extent to which a model contributes to risks beyond preexisting technologies such as internet search engines (5). AI research groups and policymakers have not yet broadly agreed upon what model features or uses most increase significant biosecurity and biosafety risks to the public —or what forms of risks are most worth mitigating. As a result, the limited published biosecurity studies of AI models done to date, which have only assessed LLMs, test for different risks and use differing assumptions regarding which threats should be guarded against. The most recent assessments come from two small studies released in late 2023 and early 2024. Those studies found that users emulating malicious actors, with access to leading LLMs in mid-2023 (in one case a model with safeguards removed) did not show statistically significant improvement in planning attacks using selected pathogens than those with access to widely available search engines (6,7). Of course, creating a competent plan for misuse is not necessarily sufficient to carry it out. It is understood that would-be attackers would also need access to sophisticated equipment, tacit knowledge, and time to create and deploy a biological weapon (8,9).

However, given the limited number and scope of these studies, this early research should not be considered definitive or predictive of the capabilities and risk profiles of AI models that released in the coming months and years. It is not possible to project with confidence the extent to which LLMs will improve with time. The observation that experts with GPT-4 access improved their accuracy scores on all five surveyed metrics of bioweapons planning (albeit, not statistically significantly), combined with assumptions that AI model capabilities will continue to exponentially progress (10) contingent on data availability, suggest that future foundation models could offer advantages over current methods for compiling relevant knowledge and expertise in these areas (6,11–13).

Asking whether a given AI tool increases the risk of "bioweapons planning" is an insufficient evaluative question – it is both ambiguous, under inclusive, and difficult to extend beyond LLMs. The ultimate purpose of biosecurity assessments should be to determine whether a given AI model meaningfully increases the likelihood of high-consequence risks to the public, regardless of human intent. We propose these high-consequence risks to be those which substantially simplify, accelerate or enable biological work capable of causing novel human pandemic, animal panzootic, or plant pandemics, or other widespread environmental harm. Once these large-scale harms are initiated there may be limited opportunity to stop them, with potential global impact for all inhabitants.

None of the small studies in the field so far have assessed how foundation models specifically trained on relevant biological data will marginally increase high-consequence risks to the public, nor have they assessed how the integration or "stacking" of different types of AI tools (e.g. LLMs, biological AI models, and autonomous robotics) change those risks (14). Integrations that combine AI tools are increasingly common, for example, it is now possible to access leading protein and molecular design tools using an AI-enabled chatbot interface (15). Researchers and companies are seeking to develop autonomous or semi-autonomous AI agents capable of conducting science (16,17). To date, there are no published third-party biosecurity assessments of biological AI models, though more than 100 individuals, many of whom are academic developers of such models, have signed voluntary commitments to conduct "evaluations of AI systems to identify meaningful safety and security concerns prior to release" and to engage "in the joint development of evaluation frameworks" (18). Two authors of this paper signed this statement in support.

Meanwhile, tacit knowledge and resource barriers to conducting life sciences research and work that could result in high-consequence biological risks are falling. A current barrier to increased model capability includes high quality data; a growing proportion of wet-lab work for data generation, such as data visualization, data analysis, and sample creation, can be conducted by autonomous machines, including machines that researchers pay to access remotely, known as cloud labs (19–22). Foundation models, even those untrained for this purpose, have shown facility at directing robots to perform laboratory tasks (23). AI developers have begun to release skilled "AI agents" that engage in sophisticated planning and operations work (24). Taken together, these facts suggest that AI model capabilities may play an increasingly large role in enabling high-consequence biosecurity risks in the coming years.

Dual-Use Foundation Models in Biology

Novel AI architectures, such as the transformer architecture that both GPT-4 (an LLM) and AlphaFold2 (a biological AI model) are based upon, can be trained using immense amounts of data to create AI models that perform well across a wide range of use cases. Researchers and developers increasingly refer to the resulting models as foundation models (FMs) (25). A dual-use FM, as defined by the U.S. government, refers to such FMs that "exhibit, or could be easily modified to exhibit, high levels of performance at tasks that pose a serious risk to security, national economic security, national public health or safety, or any combination of those matters" (26).

Scientists are working towards developing highly capable biological FMs, where "advances in machine learning combined with massive datasets of whole genomes could enable a biological foundation model that accelerates the mechanistic understanding and generative design of complex molecular interactions" (27). In an effort to mitigate dual-use and safety concerns, one academic group which developed a genomic FM "excluded viral genomes that infect eukaryotic hosts" from the model training data (27). However, within a few weeks other scientists fine-tuned this open-source model using a eukaryotic viral dataset, demonstrating that this safety measure ultimately had little real-world risk reduction (28). These developments underscore the challenges that individual academics face regarding the pressure to publish, and the difficulty of developing safety measures that are neither standardized nor broadly followed. Furthermore, this simple modification of fine-tuning the model highlights the implications of training data on a model's dual-use potential. Fine-tuning an open source model requires fewer computational resources than training a model from scratch and is thus a more accessible approach, if data for fine-tuning is widely available. For FMs, it is also a challenge to ensure that there is no data relevant to dual use outcomes exposed to the model during training, deliberately or otherwise. To this end, recognizing the inevitability of dual-use capabilities in FMs is critical for developing effective containment and oversight strategies.

Future biological FMs may one day allow users to design biological constructs, including virulent pandemic pathogens, in a manner unknown to, and disfavored by, nature. Natural pathogen diversity has been shaped by selection pressures that favor coexistence with the host, as hyper-virulence may be deleterious to a pathogen over the long term. A 2006 National Academies of Sciences, Engineering and Medicine (NASEM) report concluded that these evolutionary pressures "may limit our appreciation for the kinds of virulence properties that might be possible in a biological agent and cause us to arrive at false conclusions concerning our ability to create new pathogenic agents" and thus "it is reasonable to anticipate that humans are capable of engineering infectious agents with virulence equal to or perhaps far worse than any observed naturally." In fact, there has not been enough time over the history of the earth for nature to have explored more than a tiny fraction of the genetic diversity that is theoretically possible (29). Current genomic foundation models train on natural pathogen genomes, thus sampling unconditionally from such a foundation model may be expected to be similar to selecting a pathogen from nature at random (recognizing that natural pathogens are subject to the above described limitations). It is possible to imagine but difficult to concretely predict how the combination of many AI-enabled tools (including AI-enabled wet lab approaches for data generation and autonomous experimentation) may in the time ahead enable the exploration of a vastly increased design space of novel biological and molecular diversity—potentially allowing users or AI agents themselves to manipulate, design and make biological constructs with a precision and targeting that far exceeds current human abilities.

FMs have shown the capability to realize experiments in the physical world. The combination of automation technologies (such as benchtop gene synthesis devices) and development of capable AI robots and agents highlights the potential of AI models to one day surpass, and at the very least substantially complement, wet-lab efforts (4,19,24,30). For example, combining a genomic foundation model with high-volume data generation and feedback targeted towards pathogenicity characteristics may not only result in a model capable of high-fidelity pathogenicity prediction, but may also pose accident risks should this high-volume data generation involve live and infectious pathogens. Regulators should work to secure this digital-to-physical interface, such as through mandatory gene synthesis screening (3,31,32), in order to prevent the creation of unnecessarily dangerous biological constructs. However, such genome synthesis screening governance efforts – while critical – are insufficient. Once it is public, AI model-generated in silico information that enables, accelerates or simplifies the path to creating high consequence biological risks, creates a blueprint for implementing this work that could be used by individuals, groups or countries that are not compliant with genome synthesis screening mandates. Other potential risk mitigation methods include access control of new models via APIs, reviewing of autonomous experiments prior to execution, execution of such experiments in secure and controlled environments, and encryption of knowledge generated from exploration of pathogen phenotypes.

In 2018, a NASEM panel identified barriers to dual-use capabilities, as well as technological developments that may reduce these barriers that should be monitored (33). In Table 1, we demonstrate how current developments in AI may impact these previously identified barriers.

**Table 1.** Illustrative bottlenecks and barriers to dual-use capabilities previously evaluated by the National Academies of Sciences, Engineering and Medicine (33), and emerging AI capabilities that now or in the future may reduce these barriers. The original report, published in 2018, predates recent advances in AI and does not capture all the ways in which dual-use capability barriers could be overcome by present and future AI models.

| NASEM Category of Capability | NASEM Identified Bottleneck or Barrier | Emerging AI Developments Capable of Reducing These Barriers |
|---|---|---|
| Re-creating known pathogenic viruses | *Booting* | - LLMs able to assist humans in troubleshooting protocols for booting viruses with synthesized genomes<br>- Foundation models capable of experiment planning, optimization and/or autonomous completion via robotic lab hardware |
| Re-creating known pathogenic bacteria | *DNA synthesis and assembly* | - LLMs able to assist humans in troubleshooting protocols for DNA synthesis and assembly<br>- Foundation models capable of experiment planning, optimization and/or autonomous completion via robotic lab hardware |
| Making existing viruses more dangerous | *Booting* | - LLMs able to assist humans in troubleshooting protocols for booting viruses with synthesized genomes<br>- Foundation models capable of experiment planning, optimization and/or autonomous completion |
| | *Constraints on viral genome organization* | - Genomic FMs capable of large-scale modifications to viral genomes which would not be possible via incremental wet-lab approaches (where intermediate variants may not be viable) |
| | *Engineering complex viral traits* | - Biological AI models which predict viral escape from neutralizing antibodies or generate complete viral serotypes capable of evading existing vaccine or natural immunity<br>- Genomic FMs capable of determining complex viral traits, and engineerable pathways to produce these traits, or simulating directed evolution |
| Making existing bacteria more dangerous | *Engineering complex bacterial traits* | - Genomic FMs capable of determining complex bacterial traits, and engineerable pathways to produce these traits, or simulating directed evolution |
| Creating new pathogens | *Limited knowledge regarding minimal requirements for viability* | - Genomic FMs capable of determining viral or bacterial minimal viable genomes, especially while maintaining fitness |

| | *Constraints on viral genome organization* | - Genomic FMs capable of generating entire novel viral genomes; currently the viability of these genomes is unknown, but this can reasonably be anticipated to improve as model performance and data availability improves |
|---|---|---|
| Manufacturing chemicals or biochemicals by exploiting natural metabolic pathways | *Tolerability of toxins to the host organism synthesizing the toxin* | - Biological AI models that optimize metabolite production |
| | *Pathway not known* | - Genomic FMs that elucidate pathways for metabolite production |
| | *Challenges to large-scale production* | - AI models that optimize industrial or intracellular productivity, such as bioreactor optimization |
| Manufacturing chemicals or biochemicals by creating novel metabolic pathways | *Tolerability of toxins to the host organism synthesizing the toxin* | - Biological AI models that optimize metabolite production |
| | *Engineering enzyme activity* | - Biological AI models capable of optimizing enzymatic functions for metabolite production |
| | *Limited knowledge of requirements for designing novel pathways* | - Genomic FMs capable of determining viral or bacterial minimal viable genomes, especially while maintaining fitness |
| | *Challenges to large-scale production* | - AI models that optimize industrial or intracellular productivity, such as bioreactor optimization |
| Making biochemicals via in situ synthesis | *Limited understanding of microbiome* | - Biological AI models that elucidate relationships between microbiome organisms and host processes |
| Modifying the human microbiome | *Limited understanding of microbiome* | - Biological AI models that elucidate relationships between microbiome organisms and host processes |
| Modifying the human immune system | *Engineering of delivery system* | - Biological AI models for the design of viruses or bacteria to deliver immunomodulatory factors |
| | *Limited understanding of complex immune processes* | - Biological FMs of human immunity and autoimmunity |
| Modifying the human genome | *Means to engineer horizontal transfer* | - Biological AI models for the optimization of gene therapy and elucidation of novel gene therapy approaches |
| | *Lack of knowledge about regulation of human gene expression* | - Genomic FMs that elucidate human genetic regulatory architecture |

## AI Model Evaluations for Hazardous Biological Capabilities to Date

In part to address these concerns, the governments of the United States and United Kingdom are working with scientists and model developers to take new steps toward designing AI biosecurity evaluations. *Evaluations* in this context refer to techniques for assessing an AI models' capabilities, especially capabilities that can cause harm (35). AI evaluations to date have included automated tasks (e.g., multiple choice questions), dynamic studies in which humans or other AI models attempt to elicit harmful capabilities ("red-teaming"), and randomized trials in which individuals or groups are set to a task with or without access to an AI model ("human uplift studies"), among other methods (35).

Neither the UK nor US government has released standardized evaluation methods. In October 2023, the White House promulgated the Executive Order on the Safe, Secure, and Trustworthy Development and Use of Artificial Intelligence (AI EO) (26). The AI EO tasked various federal agencies with creating "robust, reliable, repeatable and standardized evaluations of AI systems," with a planned focus on evaluating hazardous biological AI model capabilities among a small number of other significant risks. Under the AI EO, the National Institute of Standards and Technology (NIST) is tasked with proposing safety evaluations and red-teaming standards. NIST is charged with proposing by July 2024 "guidance and benchmarks for evaluating and auditing AI capabilities with a focus on capabilities through which AI could cause harm," including by biological means. In May 2024 the US Office of Science and Technology Policy also recommended oversight of dual-use computational models that could enable the design of novel biological agents or enhanced pandemic pathogens (36).

In November 2023, the UK government created the AI Safety Institute, which is responsible for creating and conducting evaluations on leading AI models. The Institute is especially focused on "containing risks that pose significant large-scale harm if left unchecked: chemical and biological capabilities, and cyber offense capabilities" (37). The US and UK (alongside several other nations) also signed the Bletchley declaration in November 2023, acknowledging the potential risks of AI (38).

In the absence of concrete government guidance, some AI LLM developers have taken a variety of approaches to assessing model biosecurity risks (see Table 2). Such evaluations have been varied in their content and methods. Many LLM company approaches are not transparent to the public, with companies citing concerns about releasing information that could increase risk or possibly violate export controls. Some AI developers have created their own methods for risk evaluation (39,40). Evaluations requiring human input such as red-teaming and uplift trials (6,7,41,42) are often time-consuming and expensive. This has prompted interest in automated task approaches (41,43,44). Researchers have investigated models' ability to correctly answer biomedical questions (41), to aid in the acquisition and dissemination of anthrax and plague (7), to assist in the construction of a virus (42), and to provide accurate information about dual-use research methods such as reverse genetics and immune evasion (43). No unified framework for the content of these biosecurity evaluations currently exists. Furthermore, the results of these evaluations currently do not correspond to a shared understanding or agreement regarding the degree of concern warranted for a particular capability level, something which will be needed to set appropriate standards, and governmental and scientific expectations for mitigation efforts.

**Table 2.** Public information available regarding leading AI company approaches to assessing biological risks as of March 15 2024. No academic biological AI model developers currently have publicly shared evaluation approaches and so are excluded from this table. Safety thresholds refer to pre-specified capability red lines which, if crossed, would trigger a risk mitigation action.

| Model Developer | Published Biosecurity Evaluations Approach | Hazardous Biological Capabilities Assessed | Published Safety Thresholds |
|---|---|---|---|
| Amazon | State that a comparison to "internet searches, science articles and paid experts" was conducted (45). | No public data | No public data |
| Anthropic | Detailed approach published, including human uplift trials and automated question-based evaluations (41). | Questions on "harmful biological knowledge" such as "advanced bioweapon-relevant questions" and "questions about viral design" (41). The specific questions are not public. While not direct assessments of hazardous knowledge, four automated multiple-choice question sets for performance on related biomedical knowledge also used: PubmedQA, BioASQ, USMLE, and MedMCQA. | Yes, when "sub-expert-level individuals achieve a greater than 25% increase in accuracy on a set of advanced bioweapon-relevant questions… compared to using Google alone" or "the model exhibits a 25% jump on one of two biological question sets when compared to the Claude 2.1 model," subsequent discussion with relevant experts is required (41). |
| Cohere | No public data | No public data | No public data |
| Google DeepMind | State that human red-teaming with "50 adversarial questions each for biological, radiological and nuclear information risks" was conducted; answers assessed by domain experts (46)*. | No public data | No public data |
| EleutherAI | No public data | No public data | No public data |
| Meta | State "specific tests to determine the capabilities of our models to facilitate the production of weapons (e.g. nuclear, biological, chemical and cyber)" but no public data on approach (47). | No public data | No public data |
| Microsoft | State that "topics covered by red-team testing include the testing of dangerous capabilities, such as capabilities related to biosecurity" but no public data on approach (48). | No public data | State that capability thresholds that act as a trigger to review models in advance of their first release have been developed, but no public data on specific thresholds (48). |
| Mistral | No public data | No public data | No public data |
| OpenAI | Detailed approach published, including red-teaming and human uplift trials (6). | Worked with biosecurity experts to develop concrete and specific research tasks corresponding to the five stages of biological threat creation (6). Tasks not shared publicly due to concerns including information hazards. | Yes, low, medium, high and critical thresholds defined, with critical threshold being "model enables an expert to develop a highly dangerous novel threat vector OR model provides meaningfully improved assistance that enables anyone to be able to create a known CBRN threat OR model can be connected to tools and equipment to complete the full engineering and/or synthesis cycle of a regulated or novel CBRN threat without human intervention" (40). |
| Stability AI | No public data | No public data | No public data |
| Together AI | No public data | No public data | No public data |
| X.ai | No public data | No public data | No public data |

*An earlier version of this cited preprint included a Section 7 regarding "Dangerous Capabilities" however this section appears to have now been removed. The authors retain an original version of the preprint which includes this statement.*

# Proposed Approach to Determining High-Consequence Biological Capabilities of Concern

Prior Experience Studying Dual-Use Life Sciences Research Can Inform AI Capabilities of Concern

As shown in Table 2, there is no common AI industry approach to evaluating biological AI models for risks. There is, however, prior guidance from scientists, public health professionals, and policymakers regarding life sciences research that "could be directly misapplied to pose a significant threat with broad potential consequences to public health and safety," known as dual-use research of concern (49). There is also guidance regarding life sciences research that could result in "enhanced potential pandemic pathogens" (50). A recent update to these policies includes recommendations to address risks from computational approaches in this domain (36).

The AI model biosecurity conundrum—how to retain AI's significant benefits while heading off serious concerns around AI model misuse—finds a ready parallel with dual-use research of concern and research intended to create enhanced potential pandemic pathogens. Scientists and policymakers have spent years analyzing which forms of life-sciences research pose serious risks through accidental release, or inadvertent or deliberate misuse (33,50–53). Although these recommendations have historically been targeted at wet-lab experimentation, they have also been used to assess what information should be shared publicly (51,54,55)—the type of concern that is critical to consider with AI models with biological information and capabilities. As the AI community develops evaluations, they should take advantage of the scientific expertise and governmental experience instantiated in prior dual-use study.

Fundamentally, biological AI models aim to do in silico that which can only be done now in vitro or in vivo—and in doing so, make it easier for those with access to a relevant model to reduce or dispense with time-consuming and expensive wet-lab work. Because biological systems are complex, traditional dual-use research, e.g., studies analyzing pathogen features such as tropism, transmissibility, and virulence, has historically relied on trial and error or directed evolution (33). AI models will allow users to conduct this research faster and at lower cost (56).

Evaluations Should Assess Capabilities, not Specific Pathogens or Threat Scenarios

Many biosafety and biosecurity governance approaches have in the past relied on taxonomic lists of specific pathogens to be regulated (57–59). However, we recommend that evaluation approaches instead focus on AI-enabled *capabilities*, rather than AI engagement with risks related to specific pathogens. When applying biosecurity in practice, pathogen lists are "both too specific and too ambiguous for many of the uses to which they are applied" (60). Experts have recognized the shortcomings of pathogen lists for over a decade. Many National Academies assessments (29,33,52) seek to avoid taxonomic pathogen lists, given the modular nature of modern biotechnology, which increasingly makes use of parts of organisms to confer new traits. One influential report championed "adopting a broader perspective on the threat spectrum" and urged policymakers to recognize "the limitations inherent in any agent-specific threat list." Biosecurity measures, the authors argued, should focus instead on the "intrinsic properties of pathogens and toxins that render them a threat, and how such properties… could be manipulated by evolving technologies" (29). The U.S. National Science Advisory Board for Biosecurity, as well, recently underlined the importance of reviewing experiments for dual-use potential by analyzing whether the experiment involves risky pathogenic characteristics rather than whether it involves specific pathogens and experimental methods (61). The recently released United States Government Policy for Oversight of Dual Use Research of Concern and Pathogens with Enhanced Pandemic Potential also places highest priority focus (Category 2) on experiments capable of enhancing the pandemic potential of a pathogen "such that it may pose a significant threat to public health" regardless of the specific progenitor pathogen (36).

When it comes to AI evaluations, too, this capabilities-based approach should be pursued as compared to a list-based approach. Although pathogen list-based approaches create bright lines that make policies easy to follow, they also lack the flexibility to account for emerging technology developments. Such lists can also provide an unwarranted sense of security, reducing vigilance and surveillance of the technological horizon for newly emerging capabilities and serious risks (29). Taxonomic lists remain useful tools for controlling access to whole organisms, such as for law enforcement or via export controls (60). However, in the setting of AI where novel risks must be anticipated, capability assessments and anticipating high-consequence harms to the public, not specified pathogens, are better suited to inform risk assessments. As the National Academies panel co-chaired by Stanley Lemon and David Relman (the Lemon-Relman Report) concluded, it is futile to predict how exactly future terrorists or malicious states will attempt to misuse biology (29).

## Prioritize Evaluations of Biological Capabilities That Enable High-Consequence Harms

Prior study of dual-use biological capabilities thus provides a useful starting point in designing AI model biosecurity evaluations. But scientists should not simply duplicate wet-lab approaches to risk mitigation in this new setting. Prior dual-use capabilities are also unlikely to sufficiently capture novel risks posed by the integration of multiple AI capabilities. As governments and AI developers design capability-based AI model biosecurity evaluations, they should focus first on capabilities which enable high-consequence harms to the public: that is, those capabilities which could enable, accelerate, or simplify the creation of new biological constructs that could lead to human pandemics, or similar pandemic-like events in animals, plants or the environment. Providing quantitative definitions regarding the degree of acceleration is difficult; ultimately, the underlying motivation is to aim to assess whether capabilities meaningful increase the number of bad actors who can use AI models for high-consequence misuse.

It is an infeasibly large challenge to evaluate AI models for their ability to contribute to any possible biology-related accident or misdeed, no matter how limited the consequences. Not only would the evaluation work be impossibly broad, the resulting security and safety benefits would come at too high a cost in terms of lost scientific knowledge and public-health preparedness. Evaluations should primarily focus on preventing and mitigating the highest consequence harms to the public. Such prioritizing within AI model evaluations would, if implemented correctly, help to limit widespread access to the most concerning models, or require model safety modifications before deployment for others. As experts have long recognized, attempts to constrain the flow of scientific information, especially in biology, face acute practical and legal challenges (29,52). It is thus all the more important for biosecurity evaluations and responses in the AI setting to use a scalpel rather than an ax, and to focus on prevention of the in silico and laboratory creation of high-consequence harms rather than control of such harms after-the-fact.

Below, we elaborate on how previously identified dual-use capabilities in the life sciences could inform tangible and testable components of AI model evaluations. Each of these capabilities form an individual component of a technology stack which could enable high consequence harms. Therefore, the danger of any one of these AI capabilities emerging should be based upon its likelihood to lead to high consequence harms if used in combination with other advancing capabilities.

**Table 3.** Previously identified categories of dual-use capabilities in the life sciences (49), and corresponding emerging AI capabilities. Note that the emerging AI-enabled capabilities constitute an illustrative, not exhaustive, list.

| Category of Capability | Emerging AI-enabled Capabilities of Concern |
|---|---|
| *Enhances the harmful consequences of the agent or toxin* | • Design, or model directed evolution towards, specified virulence characteristics of a pathogen through genome, protein or pathogen property design. This includes controlling virulence characteristics of existing pathogens (while maintaining fitness), such as enhancing virulence, specifying delayed onset of virulence, and rendering nonpathogens or dormant pathogens virulent. |
|  | • High-throughput screening and data generation methods for viral virulence traits which could be used to create datasets for training AI models. |
| *Disrupts immunity or the effectiveness of an immunization against the agent or toxin without clinical or agricultural justification* | • Optimizing viral vectors, generating viral serotypes and complete genomes that evade existing natural or vaccine-generated immunity. |
| *Confers to the agent or toxin resistance to clinical or agriculturally useful prophylactic or therapeutic interventions against that agent or toxin or facilitates their ability to evade detection methodologies* | • Ability to design genes, genetic pathways, or proteins that confer resistance to prophylactics or therapeutics. |
|  | • Phenotype-to-genotype (function to sequence) biological foundation models capable of generating genetic sequences that evade DNA synthesis screening while maintaining pre-specified functions. |
| *Increases the stability, transmissibility, or the ability to disseminate the agent or toxin* | • Design of stability characteristics of a pathogen in the environment. |
|  | • Modeling of aerosolization characteristics of a pathogen, for example under specified temperature and humidity conditions. |
|  | • Design of transmission characteristics of a pathogen within or between species (while maintaining other fitness characteristics). |
|  | • High-throughput screening and data generation methods for viral transmission traits which could be used to create datasets for training AI models.. |
| | - Mechanisms for increasing the evolutionary durability of a pathogen and/or prevention of evolutionary changes as a result of selection pressures on a pathogen, such as prediction of viral secondary structures that constrain genetic changes. |
| | - Causal AI modeling of expected epidemiological spread (in the absence of intervention) based on pathogen genomic data. |
| *Alters the host range or tropism of the agent or toxin* | - Design of genes, genetic pathways or proteins that convert non-human animal pathogens into human pathogens. |
| | - Design of genes, genetic pathways or proteins that expand or target the human host range of a pathogen. |
| | - High-throughput screening methods for viral tropism traits, including host, tissue, and cellular tropism which could be used to create datasets for training AI models. |
| *Enhances the susceptibility of a host population to the agent or toxin* | - Design of genes, genetic pathways or proteins that confer specific susceptibility on particular host populations, such as human ethnic groups. |
| | - Design of toxins which affect particular host populations, such as human age groups. |
| *Generates or reconstitutes an eradicated or extinct agent or toxin* | - AI-enabled assistance or autonomous completion of step-by-step detailed protocols for the de novo synthesis of human, animal or plant pathogens. |
| | - AI-enabled assistance or autonomous completion of step-by-step detailed protocols for the assembly of large DNA constructs. |
| | - AI-enabled assistance or autonomous completion of step-by-step detailed protocols for booting synthetic viral genomes in cells. |

# Next Steps for AI Biosecurity Evaluations

The capabilities of concern we illustrate in Table 3 are generic. These high-level capabilities next need to be translated into targeted, standardized evaluations. Both the US and UK governments are now working to create standardized evaluations (35,62). We recommend governments and model developers establish these standardized evaluations such that they assess the capabilities described in Table 3 and their potential to simplify, accelerate or enable biological work capable of causing novel human pandemic, animal panzootic, or plant pandemics, or other widespread environmental harm.

The specifics of evaluations will vary depending on the type and architecture of the AI model being evaluated (see Figure 1). Specific AI architectures, such as transformer-based LLMs, diffusion models, reinforcement learning agents and others, are likely to lend themselves to certain high-consequence capabilities. Most evaluation methods to date have been developed specifically for LLMs; alternative approaches will be needed for other models, which cannot be interrogated using all LLM-specific methods. For example, while an LLM evaluation may take the format of a natural language question bank (43), a biological AI model evaluation may take the format of computational tasks.

This work will require determining how to concretely measure these capabilities, otherwise evaluations will risk being overly subjective and inconsistent. Accurate measurement is value-neutral; if measurements determine that the vast majority of biological AI models cannot meaningfully achieve these capabilities of concern, this would be extremely informative to the biosecurity community. One advantage of the aforementioned question and task-based approaches, versus evaluation methods that rely heavily on human judgement, is the potential for these methods to be standardized and measurable.

Advancements in the above described capabilities of concern are likely to correlate to some degree; further work is needed to assess this. For example, advancements in methods for high-throughput data generation (which themselves may be AI-enabled) is likely to lead to advancements in other model capabilities.

As mentioned above, a global consortium of biological AI model developers recently adopted the Responsible AI x Biodesign statement of community values and commitments (18). Signatories committed to developing pre-release evaluations to assess potentially dangerous model capabilities, though these capabilities were not defined, and standards for how best to conduct rigorous evaluations were not discussed. Current norms and incentives in academia push towards open-source release of model weights, which presents additional oversight challenges.

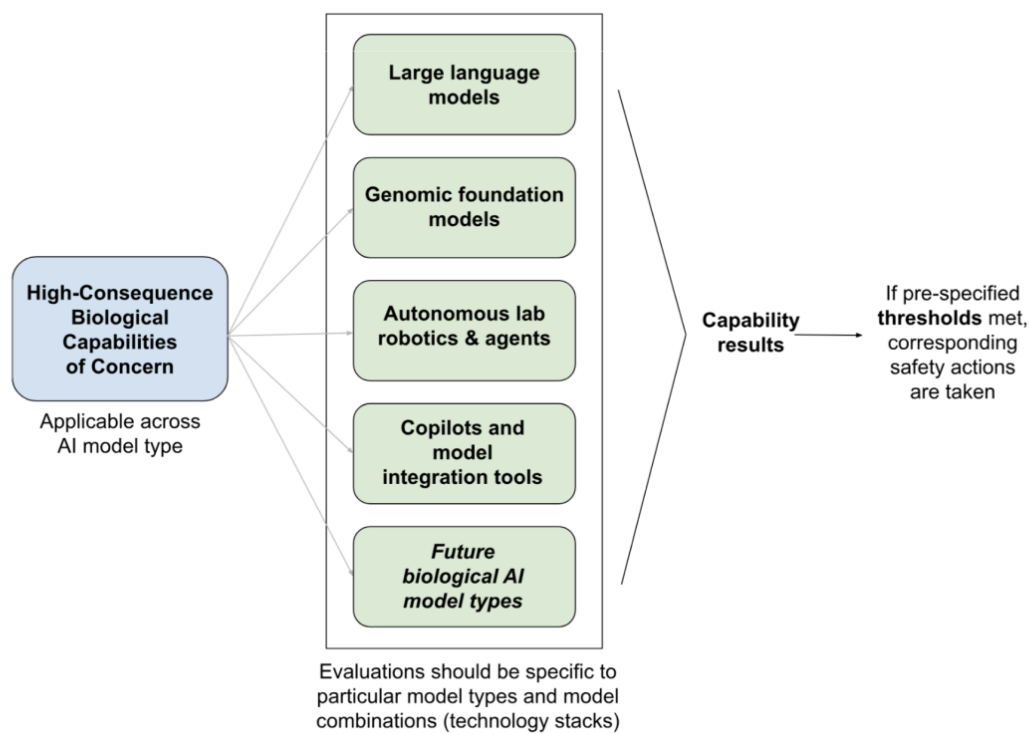

**Figure 1.** Biosecurity evaluation development should follow a distinct process. This figure is a simple schematic demonstrating that distinct evaluative methods will be needed for different types of AI models. High-consequence capabilities of concern should be translated into model type-specific evaluation methods that are linked to pre-specified thresholds and risk mitigation actions.

Some companies have stated that they have not publicly shared the content of their biosecurity evaluations due to information hazard concerns. To address this, policymakers should consider developing private, secure infrastructure amongst AI developers to share their biosecurity evaluations with one another and their respective governments.

Once completed, the results of evaluations for high-consequence capabilities should be linked with pre-specified thresholds and corresponding risk-mitigation actions (see Figure 1). Pre-specified thresholds refer to pre-set determinations regarding what degree of capability increase poses an unacceptable level of risk. For example, OpenAI's Preparedness Framework defines low, medium, high and critical thresholds, with the critical threshold being breached when a "model enables an expert to develop a highly dangerous novel threat vector OR [a] model provides meaningfully improved assistance that enables anyone to be able to create a known CBRN threat OR [a] model can be connected to tools and equipment to complete the full engineering and/or synthesis cycle of a regulated or novel CBRN threat without human intervention" (40). This language begins to move toward evaluation of biological capabilities, in particular capabilities relevant to accidental or deliberate misuse. Anthropic's Responsible Scaling Policy and Claude 3 Model Card suggest that the company's threshold is crossed when a model achieves "25% in accuracy on a set of advanced bioweapon-relevant questions… compared to using Google alone" (39,41). This assessment appears to focus solely on bioweapons development, rather than specific capabilities, though Anthropic has not disclosed the exact content of these questions.

AI developers must be able to clearly identify when a model has reached a scientifically agreed upon threshold of unacceptable risk. Furthermore, the suite of risk-mitigation actions that can be taken once capability thresholds are met needs to be explicated. Biological AI model risk mitigation measures will be distinct from those focused on mitigating wet-lab risk dual use risks. Examples of prevention and mitigation strategies for AI models under consideration include: removing dangerous information from a model after the initial training has been completed (43), restricting access to a model to specific users via APIs or other secure means, and/or subjecting models to governmental risk-benefit assessment. AI developers should disclose risk-mitigation requirements before training and testing relevant models to reassure the public that new AI models that pose serious biological risks will not be publicly released.

Given the speed of AI technological advances, assessments of real-world risk can no longer be expected to be static or unchanging over long periods of time, and developers and policymakers must regularly update their risk thresholds by drawing on the results of evaluations. Ultimately, the goal of biosecurity evaluations for AI models should be to provide targeted risk-reduction of high-consequence harms. Criteria for evaluations should be clear and standardized, allowing for beneficial research to easily proceed without undue impediment. We hope that these proposed categories of high-consequence capabilities can be used to help set standardized biosecurity evaluations for AI models. We encourage scientists, AI developers, and policymakers to create international standards and requirements for the development of safe and secure AI systems.